**DTIP** of MEMS & MOEMS 9-11 April 2008# Simulation of Coating -Visco-Elastic liquid in the Mico-Nip of Metering Size Press

Haifa El-Sadi* and Nabil Esmail
Department of Mechanical Engineering and Industrial Engineering, Concordia University
1515 St. Catherine W.,
Québec, Canada*Abstract-* **for a set of operating conditions and coating color formulations, undesirable phenomena like color spitting and coating ribs may be triggered in the Micro-nip during the coating process. Therefore, our interest in this work focus on another parameter affect on the undesirable phenomena as the vortices in the Micro-nip. The problem deals with the flow through the Micro-nip of metering size press. The flow enters and exits at a tangential velocity of 20 m/s between two rollers with diameter 80 cm and 60 μm apart. In the upper and bottom part of the domain the angular velocity is 314 rad /s. It has one sub-domain. Previous studies focus on the Micro-nip without considering the inertia and the viscoelasticity of the material. Roll coating is a technique commonly used in the coating industry to meter a thin fluid film on a moving substrate. During the film formation, the fluid is subjected to very high shear and extensional rates over a very short period of time. The fluid domain changes as a function of the hydrodynamic pressure within the nip as a result of the deformable cover usually used on one of the rolls. The free surface also adds more complexity to the flow due to the force equilibrium in the fluid gas interface. Last of all, the rheological behavior of the coating fluid is usually non-Newtonian, so the metering flow hydrodynamics is finally very difficult to describe. It is concluded that the normal forces of micro-nip increases with increasing the inhibitors. Therefore, it affects on the smoothness and creates defects. On the other hand, it can be concluded that the creation of big vortex in the middle of micro-nip affects on the coating liquid behavior.**I. INTRODUCTION

Last few years, researchers have used the lubrication theory as well as Navier Stokes equations (CFD models) to investigate reverse roll coating flows [1-7]. For the lubrication theory, the main hypothesis is that inertia effects are minimal and therefore neglected. The flow can be represented by dP/dx = μ (δ2υx/δy2) [1].

Greener and Middleman [1], using both vanishing pressure and pressure gradient as boundary conditions, predicted the coating thickness on the transfer roll for a relatively narrow range of metering rod to transfer roll speed, although the flow rate deviated from the predictions of the model due to some recirculation upstream from the nip.

Coyle et al. [3] showed that the metered nip flow deviates from the lubrication theory predictions at high speed ratios and capillary numbers. The dynamic wetting line moves towards the nip center, the nip length shrinks, and the film thickness passes through a minimum.

Most of previous studies of forward roll coating focused on the simple case of two rigid rolls separated by a narrow gap [8-10]. In many cases, one of the rolls is rubber covered, and both rolls are pressed against each other by an external load W. The coating thickness is thus governed by the operating parameters (speeds and loading), mechanical properties of the solids and the mechanical properties of the fluid. The general features of roll coating operations involving a deformable roll have been described by Coyle [115].

TABLE I
Material properties

| Material properties | α | viscosity | Relaxation time |
|---|---|---|---|
| Coating color | 0.0013 | 0.15 | 0.01 |
| Coating with 0.6% of inhibitors | 8.91 | 0.3 | 0.01 |
| Coating with 1% of inhibitors | 15.7 | 0.7 | 0.01 |

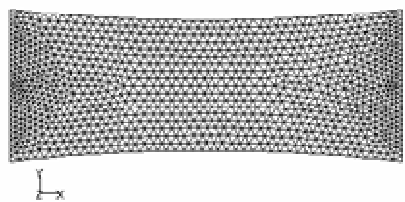

Fig. 1. mesh computational mesh generation .

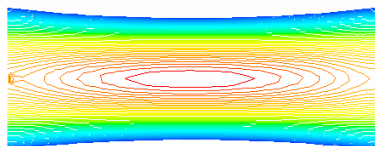

Fig. 2. the velocity contour in the micro-nip

II. NUMERICAL ANALYSIS

*A. Simulation*

Our interest is in studying the coating-visco-elastic flow in Micro- nip. Curve Fitting of Experimental Results, POLYMAT preprocessor is used for material property analysis based on experimental data as shown in table1 POLYMAT calculation is material property data that is passed to POLYDATA through a material data file. Since the liquid is visco-elastic, therefore, our interest was to find out the best visco-elastic model fit the experimental data. Here we present a quantitative fitting of KBKZ model to the experimental data of coating color without inhibitor and coating colors with 0.6% and 1%. The data show that the solution is shear thinning at low shear rate. The curve fitting shows that the KBKZ is the best model can fit the experimental data. Where KBKZ model is an integral viscoelastic model. The KBKZ model provides additional accuracy by including a damping function in its constitutive equations. The KBKZ model can calculate the extra stress tensor. The flow phenomena observed with non-Newtonian fluids cannot be predicted by the classical Navier-Stokes equation. Non-Newtonian behavior has many facets. One of them is the shear rate dependence of the shear viscosity. Therefore, we need to propose a model to describe the flow behavior of non-Newtonian fluid. The CFD package POLYFLOW 3.92 (FLUENT COMPANY) is used to solve Navier-Stokes equations and the proposed model. This CFD package uses the finite element method. It enables the use of different discretization schemes and solution algorithms, together with various types of boundary conditions. As part of the same package, (a preprocessor) Gambit is used to draw the geometry and generate the required grid for the solver. An unstructured grid with triangle elements is used as shown in Figure 1. The upper and lower walls were divided into non-uniformly spaced elements using pave meshing scheme with a ratio 0.1. The steady shear viscosity for coating colors depends on the concentration of added inhibitors. The viscous and elastic moduli of coating colors increase with increasing concentration of inhibitors, which is related to strong interaction between the particles.

©EDA Publishing/DTIP 2008ISBN: 978-2-35500-006-5

Since the viscosity is increasing with increasing the yellowing inhibitor, the viscosity of the coating liquid with 1% inhibitor was 0.7 pa.s, the viscosity of coating liquid with 0.6% inhibitor was 0.3 pa.s and the viscosity of coating liquid without inhibitor was 0.15 pa.s. the relaxation time was 0.01.

*B. Mathematical Equations*

The KBKZ model provides additional accuracy by including a damping function in its constitutive equations.the KBKZ model can calculate the extra stress tensor (T) which is = T1 +T2.

Where T1 is computed from:

$$T_1 = \frac{1}{1-\theta} \int_0^\infty \sum_{i=1}^N \frac{\eta_i}{\lambda_i^2} \exp\left(\frac{s}{\lambda_i}\right) H(I_1, I_2) \left[C_t^{-1}(t-s) + \theta C_t(t-s)\right] ds \quad (1)$$

and T2 is computed from Equation:

$$T_2 = 2\eta_2 D \quad (2)$$

where D is the rate-of-deformation tensor and $\eta_2$ is the viscosity factor for the Newtonian (i.e., purely-viscous) component of the extra-stress tensor. The viscosity ratio $r_\eta$ is defined as $\eta_2 / \eta$. The relationship of $\eta_1$ and $\eta_2$ to $\eta$ is expressed by:

$$\eta_1 = (1 - r_\eta)\eta \quad (4)$$

and $$\eta_2 = r_\eta \eta \quad (5)$$

In Equation (1), i is the index of the relaxation mode and $\theta$ is a scalar parameter that controls the ratio of the normal-stress differences:

$$\frac{N_2}{N_1} = \frac{\theta}{1-\theta} \quad (6)$$

and H is the damping function.

The Papanastasiou-Scriven-Macosko (PSM) model computes H from

$$H = \frac{\alpha}{\alpha + I - 3} \quad (7)$$

where $\alpha$ is a material parameter that primarily influences the shear-thinning behavior. The default value for $\alpha$ is 1, which may be unrealistic for many fluids, due to the large possible range of this parameter. The Papanastasiou-Scriven-Macosko (PSM) has a reversible and an irreversible function type. The irreversible PSM model allows H only to decrease.

I is computed from

$$I = \beta I_1 + (1-\beta) I_2 \quad (8)$$

I 1 and I 2 are the scalar invariants of the Cauchy-Green strain tensor:

$$I_1 = \mathrm{tr}\left(C_t^{-1}\right) \quad (9)$$

$$I_2 = \mathrm{tr}(C_t) \quad (10)$$

$C_t$ = Cauchy-Green strain tensor.

t = current time

s = metric for time integrals

$\beta$ is a material parameter that influences only the elongational behavior of the material. Table F reveals the material properties of different coating colors. It can be noticed the effect of inhibitors on the properties of coating color.

*C. Results and Discussion*

Figure 2 shows the contour of velocity distribution. It shows the formation of big vortex. . Therefore, it affects on the smoothness and creates defects. On the other hand, it can be concluded that the creation of big vortex in the middle of micro-nip affects on the coating liquid behavior